\documentclass[fleqn,10pt]{wlscirep}
\usepackage[utf8]{inputenc}
\usepackage[T1]{fontenc}
\usepackage{amssymb}
\usepackage{float}
\usepackage{graphicx}
\usepackage{caption}
\usepackage{subcaption}
\usepackage{float}
\usepackage{geometry}
\usepackage{pythonhighlight}
\usepackage{multirow}
\usepackage{dirtree}
\usepackage[numbers]{natbib}
\usepackage{multirow}
\usepackage{gensymb}
\usepackage{url} 
\providecommand{\urlstyle}[1]{}

\title{A multiresolution weather dataset for the Southwestern South Atlantic (2017-2018)}

\author[1]{Luan C. V. Silva}
\author[1]{L\'{\i}via Sancho}
\author[1]{Mauricio S. Silva}
\author[3]{Elisa Passos}
\author[1]{Larissa F. R. Jacinto}
\author[1,2]{Rebeca S. Lyra}
\author[1,2]{Nilton O. Moraes}
\author[1]{Carina S. Bock}
\author[1]{Douglas M. Nehme}
\author[1]{Raquel Toste}
\author[1]{Jacques Honigbaum}
\author[1]{Rodrigo S. Luna}
\author[4]{Carlos H. Beisl}
\author[1]{Patricia M. Silva}
\author[1]{Adriano O. Vasconcelos}
\author[1]{Rian C. Ferreira}
\author[5]{C{\'e}dric Eneau}
\author[1]{Fernando A. Rochinha}
\author[1,2]{Luiz P. F. Assad}
\author[1]{Alvaro L. G. A. Coutinho}
\author[1]{Laura Bahiense}
\author[1]{Alexandre G. Evsukoff}

\affil[1]{COPPE, Universidade Federal do Rio de Janeiro}
\affil[2]{Instituto de Geoci{\^e}ncias - IGEO, Universidade Federal do Rio de Janeiro}
\affil[3]{Faculdade de Oceanografia, Universidade do Estado do Rio de Janeiro}
\affil[4]{Geospatial Petroleum}
\affil[5]{TotalEnergies OneTech}

\begin{abstract}

The Southwestern South Atlantic (SWSA) is a key region for climate research and renewable energy assessment, yet high-resolution meteorological data are scarce. We present a multiresolution dataset spanning February 2017–November 2018, combining Weather Research and Forecasting (WRF) simulations with Sentinel-1A/B Synthetic Aperture Radar (SAR) wind fields processed using the CMOD5 model. WRF outputs were generated every 30 minutes for three nested domains (9 km, 3 km, 1 km) through 975 short-term simulations. SAR/CMOD5 wind fields are provided at 500 m and 1 km resolution across 104 acquisition dates. Validation shows strong agreement: daily spatial averages of 10 m wind speed yield RMSE and MAE below 3 m/s on over 93\% of acquisition days, while more than 91.5\% of pixel-level residuals fall within ±3 m/s. In situ measurements from the Itaja{\'i} buoy further confirmed the reliability of both sources. The dataset supports regional climate studies, wind energy resource assessment, and machine-learning applications in forecasting and downscaling, with usage examples included to aid practical adoption.

\end{abstract}

\begin{document}

\flushbottom
\maketitle

\thispagestyle{empty}

\section{Background \& Summary}


The South Atlantic plays a pivotal role in the global climate system. It is the only ocean basin with a net equatorward heat transport, directly influencing heat redistribution, carbon storage, and global climate variability \citep{chidichimo2023energetic}. The limited understanding of local physical processes and climate variabilities in the Southwestern South Atlantic (SWSA) is largely attributed to sparse observational data and inaccuracies in long-term model integration \citep{gramcianinov2023recent}. 

Brazil is already among the world's largest producers of renewable electricity, with hydropower accounting for more than half of its electricity generation, leveraging the extensive river systems \citep{catolico2021socioeconomic}. However, the region also has significant potential for other sources of large-scale renewable energy generation. The coastal area of Brazil has the potential for offshore wind energy estimated at $\approx$ 3 TW and more than 14,800 TWh of average annual electricity production \citep{azevedo2020assessment, gonzalez2020regulation} and wave energy, with the potential to reduce CO\textsubscript{2} emissions by approximately 44.52 million tons per year if gas-fired thermal energy is replaced \citep{bastos2023wave}. According to Vinhoza \& Schaeffer \citep{vinhoza2021brazil}, few studies have comprehensively assessed Brazil's offshore potential, and those that exist have focused primarily on gross potential, often neglecting technological, environmental, and social constraints. While global models provide coverage of the region, increasing spatial resolution significantly enhances the representation of meteorological features such as wind variability and extreme events \citep{hadjipetrou2024high, ryu2025increasing}.

 Beyond improving our understanding of SWSA weather dynamics, a multiresolution weather dataset also provides rich training material for artificial intelligence (AI) methods. Owing to booming computing power, it is now possible to simulate climate systems at high resolution. This, combined with a surge in climate observations from weather satellites, creates a higher data volume that allows for the application of advanced machine learning methods to improve climate modelling and prediction \citep{bracco2025machine}. However, data limitations are still recognized as a fundamental challenge in weather forecasting \citep{pagano2024challenges, chen2023machine}. AI can reduce both cost and time for delivering high-resolution weather data \citep{wu2024data}. 

In this study, a dataset covering the southeastern coast of South America and the adjacent South Atlantic Ocean was created by integrating numerical simulations and satellite observations at multiple spatial resolutions. High-temporal-resolution (30 min) outputs were generated by the Weather Research and Forecasting (WRF) model \citep{skamarock_etal_2021} through 975 independent runs-each lasting approximately 5.5 hours-for a combined computation time of approximately 5,362.5 hours. SAR images from Sentinel-1A/B C-band sensors \citep{hadjipetrou2022evaluating,6504845} were processed with the CMOD5 algorithm to estimate 10 m wind fields. Wind fields from the numerical and satellite models were validated against height-corrected in situ buoy measurements. This dataset spans February 2017 to November 2018 and provides exciting possibilities for AI applications, including forecast emulation \citep{li2024generative, wang2019fast} and cross-scale downscaling \citep{oh2022machine}. The different data sources available allow observation-constrained bias correction \citep{zhang2024deep, wang2022deep}, and the  wide range of weather variables allows multivariate analysis \citep{zhu2023weather2k}, leading to improved results and better exploitation of correlations among various atmospheric parameters.

\section{Methods}

\subsection{Study area}

\begin{figure}[H]
\centering
\includegraphics[width=0.9\textwidth]{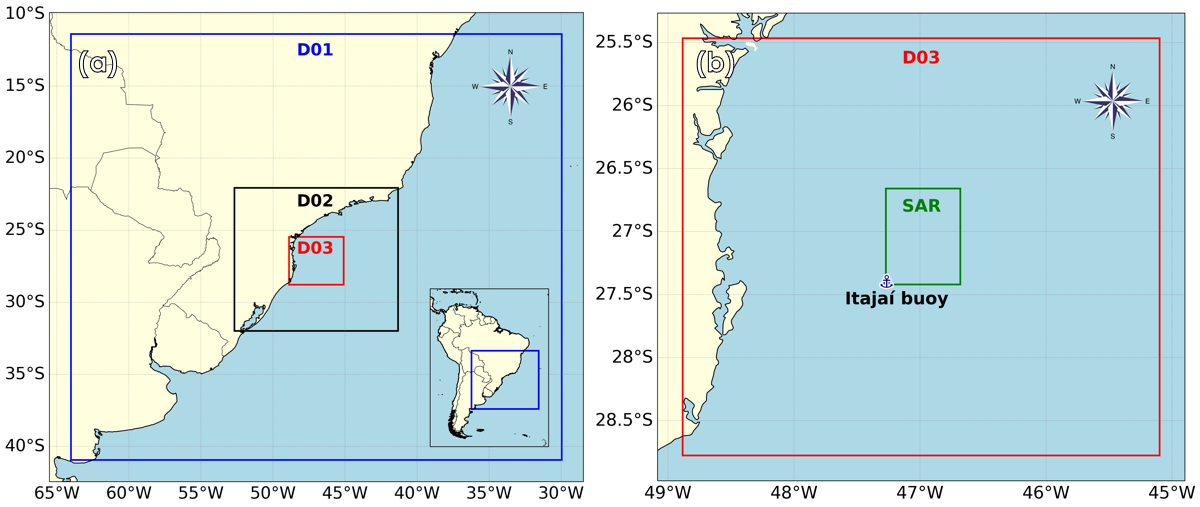}
\caption{Study area. (a) Three-nested domains configured for the Weather Research and Forecasting model (WRF); D01 with 9 km resolution in blue, D02 with 3 km resolution in black, and D03 with 1 km resolution in red. (b) WRF model's most refined grid (D03), selected SAR region and the location of the Itaja{\'i} meteo-oceanographic buoy.}\label{fig:study_area}
\end{figure}

The study area is situated in the southwestern Atlantic, with a focus on the southern coast of Brazil and its adjacent offshore regions. To ensure adequate spatial coverage and resolution across the study region, a system of nested computational domains was constructed. The simulation used a one-way nesting strategy with three domains, as shown in Figure \ref{fig:study_area}a. The outermost domain (D01), shown in blue, covers a broad area at a 9 km grid resolution; an intermediate domain (D02), in black, with a grid resolution of 3 km, is embedded within D01 and serves as a bridge by receiving input from D01 and passing boundary information to the innermost domain. The innermost domain (D03), shown in red, focuses on the core study area, which was configured with a grid resolution of 1 km (Figure \ref{fig:study_area}a). 

The area of the wind speed computed from the SAR  acquisition frames overlapping the study area, serving as a  reference for data validation along with the Itaja{\'i} meteo-oceanographic buoy location, are shown in Figure \ref{fig:study_area}b. This combination of nested atmospheric domains and satellite coverage enables a detailed examination of wind variability from regional to local scales, with particular focus on the high-resolution D03 area.

\subsection{The Weather Research and Forecasting (WRF) model}

The atmospheric simulations were conducted using version 4.3.3 of the WRF model, which was configured with the Advanced Research WRF (ARW) dynamic core. This core integrates the full set of Navier-Stokes equations, thus avoiding the hydrostatic assumption and allowing for a detailed representation of atmospheric flows. Vertical discretization employs a terrain-following hybrid sigma-pressure coordinate. The model framework adopts the Arakawa staggered C-grid \citep{Mesinger_Arakawa_1976} and uses finite differences to solve the governing equations \citep{Lundquist_etal_2010}. Staggered grids solve quantities in different regions of the grid, such as corners, centres, or faces, which means that dimensions can vary at one point depending on the variable. The mass, thermodynamic, scalar, and chemical variables are calculated in the centre of the grid cell. Moreover, the U and V components of the horizontal velocity are normal to the corresponding faces of the cell \citep{skamarock_etal_2021}. Thus, all dimensions with the suffix "stag" are located normal to the corresponding faces of the cell, while the others are located in the centre of the cell.

\begin{table}[ht]
\small
  \centering
  \caption{WRF model configuration.}
  \label{tab:wrf_conf}
  \begin{tabular}{@{} ll @{}}
    \hline
    \textbf{Description} & \textbf{Configuration} \\
    \hline
    \multicolumn{2}{@{}l}{\textbf{Timing}} \\[4pt]
    Simulation period & 06:30h 06/02/2017 to 06h 18/11/2018 \\ 
    Time step & 48s \\ 
    Output Frequency & \textcolor{black}{30 min} \\ 
    \hline
    \multicolumn{2}{@{}l}{\textbf{Maps and grids}} \\[4pt]
    Map projection & Mercator \\ 
    Horizontal grid spacing & 9, 3 and 1 km (361 x 355) \\ 
    Grid Points & 361 x 355 (all grids) \\ 
    Vertical levels & 45 \\ 
    Nesting strategy & One-way \\ 
    Topography data & GMTED2010 30 s (1 km) \citep{Danielson2011}, GMTED2010 30 s (1 km) \citep{Danielson2011} and SRTM 9 s (90 m) \citep{usgs_2023} \\ 
    Land Use and Land Cover data & MODIS 30 s (1 km)  \citep{MCD12Q2_006}, MODIS 30 s (1 km)  \citep{MCD12Q2_006}, and MAPBIOMAS 30 m \citep{MapBiomas} \\ 
    \hline
    \multicolumn{2}{@{}l}{\textbf{Forcing strategy}} \\[4pt]
    Initial and Boundary Conditions & FNL GFS (0.25 km) + SST from ROMS model on D03 \\ 
    Boundary Conditions Frequency & 6 hours \\ 
    Initialization & 00 hour \\ 
    Runs duration & 2 days + 6 hours (54 hours, 109 times)\\ 
    Spin-up time & 6 hours \\ 
    \hline
    \multicolumn{2}{@{}l}{\textbf{Physical parametrization schemes}} \\[4pt]
    Shortwave radiation & RRTMG \citep{iacono_2008} \\ 
    Longwave radiation & RRTMG \citep{iacono_2008} \\ 
    Cumulus parametrization & Multi-Scale Kain-Fritsch \citep{Zheng_2016} \\ 
    Microphysics & Morrison Double-Moment \citep{morrison_2009} \\ 
    Surface Layer & Revised MM5 scheme for the surface layer \citep{jimenez_2012} \\ 
    Land surface model & Noah-MP  \citep{niu_2011} \\ 
    Planetary boundary layer & Yonsei University (YSU)  \citep{hong_2006}\\ 
    Turbulence Diffusion Option & 2 \\ 
    \hline
  \end{tabular}
\end{table}

As a consequence, the number of grid points for the "stag" variables is one unit greater than the other variables located in the centre of the cell grid. In this way, the variables XLAT and XLONG store the coordinate values located at the centers of the grid cells, whereas the variables XLAT\_U, XLONG\_U, XLAT\_V, and XLONG\_V store the coordinate values located at the respective faces of the grid cells. Detailed information can be found on the WRF website\footnote{https://amps-backup.ucar.edu/information/configuration/wrf\_grid\_structure.html}.

Physical processes, including the radiation balance, microphysical processes, convection, turbulence, and surface interactions, are incorporated through parameterization schemes. In the vertical dimension, all the domains shared the same discretization, composed of 45 vertical levels distributed throughout the atmospheric column. Close to the surface, enhanced resolution was applied with four vertical levels approximately 200 metres above mean sea level, located at 26, 60, 103, and 160 metres, respectively. The nested domains were centred at 27.12\degree S and 46.99\degree W, using the Mercator projection. This positioning ensures dynamic consistency of the physical processes within the model and avoids inconsistencies or artificial relaxation effects at the boundaries of D03 relative to the core area of interest.

The physical options of the model were selected on the basis of previous evaluations and applications in similar contexts \citep{dragaud_2021,soares_da_silva_2023,deSouzaetal_2024}. The chosen schemes are the following: the Morrison double moment scheme for cloud microphysics \citep{morrison_2009}, the multiscale Kain--Fritsch scheme for convective processes \citep{Zheng_2016}, the rapid radiative transfer model for general circulation models (RRTMG) for both longwave and shortwave radiation \citep{iacono_2008}, the Noah-MP scheme for land surface modelling \citep{niu_2011}, the revised Mesoscale Model (MM5) scheme for the surface layer \citep{jimenez_2012}, and the Yonsei University scheme for boundary layer dynamics \citep{hong_2006}. Horizontal diffusion was treated in physical space to ensure better accuracy over complex terrain. The overall model configuration is displayed in Table \ref{tab:wrf_conf}. 

Another important change in the data concerns the precipitation variables (RAINC and RAINNC). Originally, the precipitation data produced by WRF are cumulative over the entire duration of the one run (every 54 hours in the present case); however, the precipitation values made available are cumulative only between time intervals (30 minutes). Some variables can be derived indirectly from others, as shown in Table \ref{tab:variables_formulas}, which presents a relevant subset of such relationships.

\begin{table}[ht]
\centering
\caption{Variables derived from WRF outputs and their formulas.}
\label{tab:variables_formulas}
\begin{tabular}{ll}
\hline
\textbf{Variable} & \textbf{Formula} \\
\hline
Atmospheric pressure ($P_{\text{atm}}$) & $P + PB$ \\
Geopotential height & $PH + PHB$ \\
Height above mean sea level ($z$) & $\frac{PH + PHB}{g}, \quad g = 9{,}81\,\text{m/s}^2$ \\
Potential temperature ($\theta$) & $T + 300$ \\
Air temperature ($T_{\text{air}}$) & $\theta \left( \frac{P + PB}{100} \times \frac{1}{1000} \right)^{\frac{R}{c_p}}, \quad \frac{R}{c_p} = 0{,}2854$ \\
\hline
\end{tabular}
\end{table}

The simulations utilized initial (IC) and boundary conditions (BC) from the final analysis (FNL) provided by the National Centers for Environmental Prediction (NCEP), which were derived from the Global Forecast System (GFS2). This dataset offers global atmospheric fields with 0.25\degree\ horizontal resolution (approximately 24 km) and a temporal frequency of 6 hours. The GFS model has undergone continual development since the 1980s, with the most recent upgrade documented in 2021\footnote{https://emc.ncep.noaa.gov/emc/pages/numerical\_forecast\_systems/gfs/documentation.php}. To improve the terrain representation in the boundary data, topographic information from the Shuttle Radar Topography Mission (SRTM) \citep{usgs_2023} and land cover data from the MapBiomas project \citep{MapBiomas} were incorporated. This approach follows the methodological refinements proposed in recent works \citep{Pedruzzi_2022,Jacinto_2020}. Additionally, to improve the air--sea interactions in the innermost domain (D03), sea surface temperature (SST) results from a hydrodynamic model were implemented. More details can be found in item \ref{roms}. A summary of the main features of the simulations is presented in Table \ref{tab:wrf_conf}.

\paragraph{Sea surface temperature (SST) boundary condition.}\label{roms}

The hourly SST boundary conditions, with a horizontal resolution of $\approx$1 km, were obtained from a Regional Ocean Modelling System (ROMS) simulation. ROMS is a widely used three-dimensional free-surface hydrostatic model that solves primitive equations in a discretized space using an Arakawa C grid \citep{shchepetkin_mcwilliams2005}. Hydrodynamic simulation was performed on a regional grid with 40 vertical layers and encompasses the region between 25.3\degree S and 28.9\degree S in latitude and 45.0\degree W and 49.0\degree W in longitude, i.e., including the D03 domain of the WRF model. The bathymetry was extracted from the Earth TOPOgraphy (ETOPO) 2022 \citep{macFerrin_etal2025} global model. The simulation was performed using the European Centre for Medium-Range Weather Forecasts (ECMWF) Reanalysis version 5 (ERA5) \citep{hersbach2023era5} as surface forcing and the daily results from the Copernicus Marine Environment Monitoring Service's GLORYS12V1 (Global Ocean Physics Reanalysis) product \citep{lellouche_etal2021} as the IC and lateral BC. Astronomical tides from the TOPEX/Poseidon TPXO7.2 global model were also used as forcing at open boundaries \citep{egbert_etal1994, egbert_erofeeva2002}. SST results were validated using the Operational Sea Surface Temperature and Ice Analysis dataset \citep{good_etal2020} and in situ data from the National Buoy Program (PNBOIA\footnote{https://www.marinha.mil.br/chm/dados-do-goos-brasil/pnboia}) network.

\subsection{Wind field estimation from synthetic aperture radar (SAR) images}

\paragraph{SAR imagery.}\label{sar}

The wind field was estimated using data from the Sentinel-1A and Sentinel-1B satellites, provided free of charge through the European Space Agency (ESA) Sentinel portal\footnote{https://dataspace.copernicus.eu/explore-data/data-collections/sentinel-data}. The data were acquired in Interferometric Wide Swath (IW) mode, featuring a swath width of approximately 250 km and a spatial resolution of 5 m $\times$ 20 m (ESA, 2024). The original datasets are available at Level-1 processing in ground range detected (GRD) format, which consists of focused SAR data that have been detected, geocoded, and projected onto the ground range using the WGS84 ellipsoid model. The pixel values represent radar backscatter intensity in this format, with phase information discarded. 

For the present study, a total of 104 SAR images were acquired over the offshore region of interest. The temporal availability of validated reference data from the Itaja{\'i} metoceanographic buoy guided image selection. Although the nominal temporal resolution of Sentinel-1 is 12 days per satellite, the combined coverage of Sentinel-1A and Sentinel-1B over the region of interest results in an effective revisit time of approximately 6 days for descending orbits. However, the image frames of the two satellites are spatially offset. Figure \ref{fig:study_area}b illustrates the spatial configuration of the Sentinel-1A and Sentinel-1B acquisition frames. The Sentinel-1A frame is deliberately offset relative to the Sentinel-1B frame, enabling both enhanced temporal coverage and reduced imaging gaps.

The coverage area of the Sentinel-1A frame is approximately 44,558 km$^{2}$, while that of Sentinel-1B is about 52,660 km$^{2}$. The overlapping region between the two frames spans 32,500 km$^{2}$, corresponding to 73.0\% and 61.7\% of the Sentinel-1A and Sentinel-1B frames, respectively. Notably, the Itaja{\'i} metoceanographic buoy is located near the southern edge of the Sentinel-1A frame, whereas it is positioned closer to the  of the Sentinel-1B frame. This spatial discrepancy may introduce edge effects in the wind field estimates derived from the Sentinel-1A data, potentially influencing comparisons with buoy measurements. To mitigate these effects and ensure consistency in the spatial analysis between the two acquisitions, a subset was defined on the basis of the intersection area between the Sentinel-1A and Sentinel-1B scenes (SAR layer in Figure \ref{fig:study_area}b). This subset focuses on the central portion of the overlapping region and includes the buoy location, thereby reducing spatial mismatches and allowing for a more robust comparison of wind estimates across sensors.

\begin{figure}[H]
\centering
\includegraphics[width=0.9\textwidth]{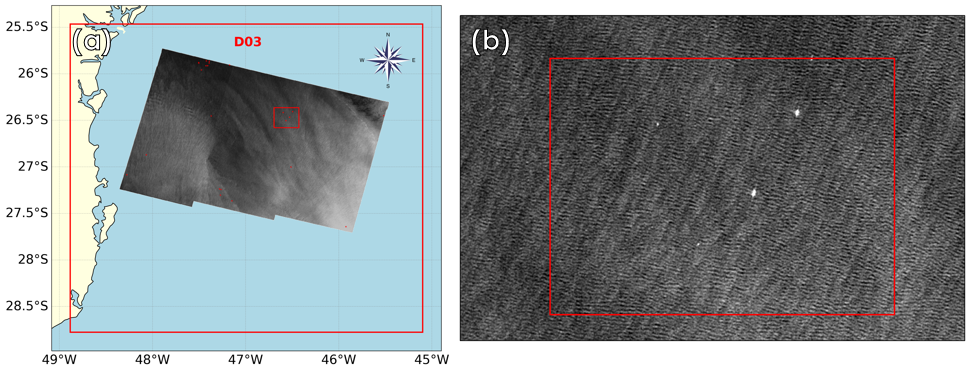}
\caption{(a) Bright object detection mask produced by the SNAP platform, highlighting potential ship targets; (b) detailed view of the corresponding objects identified in the SAR image.}\label{fig_ship}
\end{figure}

In SAR imagery, ships appear as strong backscatter and are represented as bright targets against the ocean surface (Figure \ref{fig_ship}). Masking these high-reflectivity objects prevents such targets from introducing artefacts into the wind speed retrieval process. Without this masking step, ship signatures may lead to erroneous wind speed estimates, manifesting as outliers in the derived wind field.

A series of specific functions was developed to detect and mask anomalous reflectors, thereby mitigating this issue. Ship detection is based on identifying point-like intensity peaks that significantly exceed the surrounding pixel values. In contrast, oil slick detection focuses on regions with anomalously low backscatter intensities.

The primary function implemented for this purpose, named \textit{ship}\_\textit{detection}, is designed to identify both types of anomalies. By default, pixels with values above the 99.7$^{th}$ percentile of the image intensity distribution are classified as ships, whereas pixels below the 0.01$^{st}$ percentile are flagged as potential oil slicks. These threshold values are configurable and can be adapted to suit specific application requirements.

Following anomaly detection, the \textit{dilation}\_\textit{mask} function is applied to expand the initially identified pixel regions. This dilation step is crucial for eliminating discontinuities and forming cohesive, contiguous masks. The operation uses a default square structuring element of 3 $\times$ 3 pixels, although this parameter can be adjusted depending on the desired level of mask expansion.

To reconstruct the masked regions in the SAR images, we employed an image inpainting function, a technique commonly used for restoring degraded or missing areas in digital images, such as damaged photographs. This function provides two available methods: terrain elevation-dependent attenuation (TELEA) and the Navier--Stokes method. The TELEA method uses the fast marching method to propagate known pixel values into the masked regions iteratively \citep{telea2004image}. On the other hand, the Navier--Stokes method applies partial differential equations grounded in fluid dynamics to guide the reconstruction process \citep{bertalmio2001navier}. A comparative analysis was conducted to evaluate the performance of both techniques within the context of wind field estimation from SAR images. The Navier--Stokes method was selected as the default approach because of its superior performance in maintaining image continuity and minimizing edge artefacts. Nevertheless, the inpainting method remains a configurable parameter within the implementation, allowing users to adapt the choice according to specific application requirements.

\paragraph{The CMOD5 geophysical model.}\label{cmod5}
The subsequent stage of the methodology involves applying the CMOD5 geophysical model function (GMF), developed by Hersbach in 2003 \citep{hersbach2003cmod5}. This empirical model is designed to estimate the near-surface wind speed and direction from the normalized radar cross-section (NRCS or $\sigma^0$) obtained from C-band SAR measurements. 

As wind interacts with the ocean surface, it generates small-scale roughness patterns that tend to align with the wind direction. These surface features modulate the radar backscatter signal, a relationship well documented in the literature (e.g., Wackerman et al. \citep{wackerman2003automated}). The CMOD5 model captures this relationship by expressing $\sigma^0$ as a function of three key variables: wind speed at 10 m above sea level ($u_{10}$, in m/s), the relative angle between the radar look direction and wind direction ($\Phi$, in degrees), and the local incidence angle of the radar beam ($\theta$, in degrees). Formally, the model is expressed as $\sigma^0 = f(u_{10},\Phi,\theta)$, where \(f\) is a nonlinear empirical function calibrated using extensive datasets from scatterometers, SAR observations, buoys, and numerical weather prediction models.

The wind direction is first estimated using the u and v components of the wind vectors provided by the WRF model. These components are interpolated to match the spatial resolution defined for the SAR-based analysis. The wind direction is computed for each grid cell and subsequently compared with the direction retrieved from SAR imagery using the CMOD5 model. The wind speed field is derived by applying the CMOD5 geophysical model to the Sentinel-1 SAR data. This process yields two wind speed products: one derived from the WRF model and the other obtained directly from the SAR imagery through inversion of the CMOD5 function.

\section{Data records}

The dataset acquired in this data descriptor is available at the Harvard Dataverse\footnote{https://dataverse.harvard.edu/dataset.xhtml?persistentId=doi:10.7910/DVN/OUEZ0H}. The dataset is composed of: (1) WRF model outputs with selected variables (Table \ref{tab:wrf_variables}), (2) SAR/CMOD5 wind spatial retrievals, (3) in situ buoy data (single location). Table \ref{tab:grid_sizes} shows the grid sizes of in the dataset for WRF and SAR/CMOD5 layers for the available spatial resolutions. Note that the WRF grid dimensions listed in Table \ref{tab:grid_sizes} refer to the number of mass-point (C-grid) locations in the horizontal (X-Y) plane-that is, the unstaggered model grid. These values represent the size of a single horizontal slice of mass centres in the WRF output.

\begin{table}[ht]
  \centering
  \caption{Grid dimensions for WRF and SAR/CMOD5 10 m wind speed data at different spatial resolutions.}
  \label{tab:grid_sizes}
  \begin{tabular}{lcc}
    \hline
    \textbf{Resolution} & \textbf{WRF} & \textbf{SAR/CMOD5} \\
    \hline
    500\,m   & -- & $164\times114$ \\
    1\,km    & $354\times360$   & $82 \times 57$    \\
    3\,km    & $354\times360$   & --  \\
    9\,km    & $354\times360$   & --               \\
    \hline
  \end{tabular}
\end{table}

\paragraph{WRF data}

The WRF output is provided for three nested domains (D01, D02, and D03) spanning from 00:30\,UTC 06 February 2017 through 06\,UTC 18 November 2018 at 30-minute temporal resolution. Table \ref{tab:wrf_variables} lists the WRF variables in the dataset. The first seven are coordinates, while the rest are data variables. For more details, refer to the WRF Users' Guide\footnote{https://www2.mmm.ucar.edu/wrf/users/docs/user\_guide\_v4/contents.html}. The WRF outputs were stored as daily NetCDF (\texttt{.nc}) files, with each file reaching approximately 480 GB. Separate files are available for each model grid, and all variables are stored together with their corresponding coordinates.

\begin{table}[ht]
\small
  \centering
  \caption{Description of WRF output variables available in the dataset.}
  \label{tab:wrf_variables}
  \begin{tabular}{llll}
    \hline
    \textbf{Variable} & \textbf{Description and units} & \textbf{Dimensions} \\
    \hline
    Times         & Reference date (calendar date) & 1D (Time) \\
    
    XLAT         & Latitude values of grid cell centre points (degree\_north) & 2D (south\_north, west\_east) \\
    
    XLONG         & Longitude values of grid cell centre points (degree\_east) & 2D (south\_north, west\_east) \\
    
    XLAT\_U         & Latitude values of grid cell zonal faces (degree\_north) & 2D (south\_north, west\_east\_stag) \\
    
    XLONG\_U         & Longitude values of grid cell zonal faces (degree\_east) & 2D (south\_north, west\_east\_stag) \\
    
    XLAT\_V         & Latitude values of grid cell meridional faces (degree\_north) & 2D (south\_north\_stag, west\_east) \\
    
    XLONG\_V         & Longitude values of grid cell meridional faces (degree\_east) & 2D (south\_north\_stag, west\_east) \\

    LU\_INDEX & Land use index (categorical) & 2D (south\_north, west\_east) \\
    
    HGT       & Terrain height above sea level (m) & 2D (south\_north, west\_east) \\
    
    LANDMASK  & Land-sea mask (1 = land, 0 = water) & 2D (south\_north, west\_east) \\

    ZNU       & Vertical coordinate at full (mass) levels (m) & 2D (Time, bottom\_top) \\
    ZNW       & Vertical coordinate at half (staggered) levels (m) & 2D (Time, bottom\_top\_stag) \\

    T2        & 2 m air temperature (K) & 3D (Time, south\_north, west\_east) \\
    
    TH2       & 2 m potential temperature (K) & 3D (Time, south\_north, west\_east) \\
        
    SST       & Sea surface temperature (K) & 3D (Time, south\_north, west\_east) \\

    PHB       & Base-state geopotential ($\mathrm{m}^2\!\cdot\!\mathrm{s}^{-2}$) & 3D (bottom\_top\_stag, south\_north, west\_east) \\
    PB        & Base-state pressure (Pa) & 3D (bottom\_top, south\_north, west\_east) \\

    PSFC      & Surface pressure (Pa) & 3D (Time, south\_north, west\_east) \\
    
    RAINC     & Cumulative convective precipitation ($\mathrm{kg}\!\cdot\!\mathrm{m}^{-2}$) & 3D (Time, south\_north, west\_east)  \\
    
    RAINNC    & Cumulative non-convective precipitation ($\mathrm{kg}\!\cdot\!\mathrm{m}^{-2}$) & 3D (Time, south\_north, west\_east) \\

    U10       & 10 m U-component of wind ($\mathrm{m}\!\cdot\!\mathrm{s}^{-1}$) & 3D (Time, south\_north, west\_east) \\
    
    V10       & 10 m V-component of wind ($\mathrm{m}\!\cdot\!\mathrm{s}^{-1}$) & 3D (Time, south\_north, west\_east) \\
    
    T         & Perturbation potential temperature (K) & 4D (Time, bottom\_top, south\_north, west\_east) \\
    
    PH        & Perturbation geopotential ($\mathrm{m}^2\!\cdot\!\mathrm{s}^{-2}$) & 4D (Time, bottom\_top\_stag, south\_north, west\_east) \\
        
    P         & Perturbation pressure (Pa) & 4D (Time, bottom\_top, south\_north, west\_east) \\
            
    U         & U-component of wind ($\mathrm{m}\!\cdot\!\mathrm{s}^{-1}$) & 4D (Time, bottom\_top, south\_north, west\_east\_stag) \\
    
    V         & V-component of wind ($\mathrm{m}\!\cdot\!\mathrm{s}^{-1}$) & 4D (Time, bottom\_top, south\_north\_stag, west\_east) \\
    
    W         & Vertical velocity ($\mathrm{m}\!\cdot\!\mathrm{s}^{-1}$) & 4D (Time, bottom\_top\_stag, south\_north, west\_east) \\   

    \hline
  \end{tabular}
\end{table}
 
WRF output is provided for three nested domains (D01, D02, D03) spanning from 00:30\,UTC 06 February 2017 through 06\,UTC 18 November 2018 at 30-minute temporal resolution. Table \ref{tab:wrf_variables} lists the WRF variables in the dataset. The first seven are coordinates, while the rest are data variables. For more details, refer to the WRF Users' Guide\footnote{https://www2.mmm.ucar.edu/wrf/users/docs/user\_guide\_v4/contents.html}. The WRF outputs were stored as daily NetCDF (\texttt{.nc}) files, with each file reaching approximately 480 GB. Separate files are available for each model grid, and all variables are stored together with their corresponding coordinates.

\begin{figure}[H]
\centering
\includegraphics[width=0.9\textwidth]{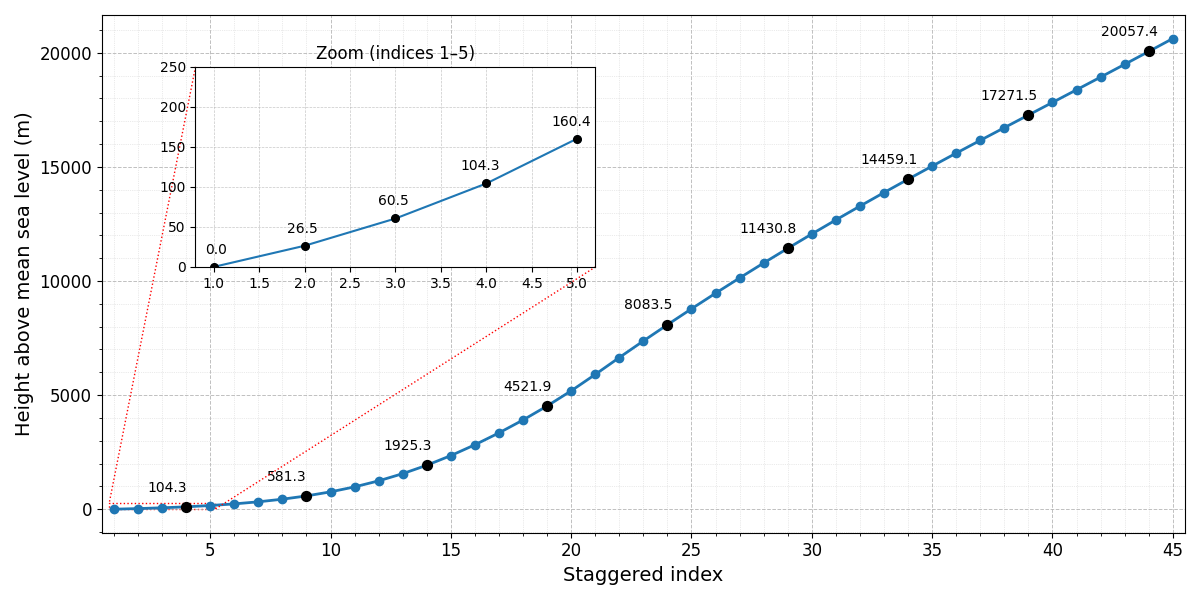}
\caption{Approximate (pressure-dependent) height above mean sea level for each WRF-staggered vertical index, computed as 
\((PH + PHB)/g\), where \(PH\) and \(PHB\) are are the perturbation and base-state geopotential respectively, and \(g\) is the gravitational acceleration.}\label{fig:heights}
\end{figure}

Bilgili et al. (2022) reported that the 2020 mean hub height of offshore wind turbines was 104.03 m \citep{bilgili2023potential}. Figure \ref{fig:heights} shows the corresponding heights at a range of vertical levels, calculated using the formula given in Table \ref{tab:variables_formulas}, with the highest levels exceeding 2,000 m. However, the full set of "4D" variables required to represent every model level produces very large files. Many climate-related applications, such as assessments of offshore wind-energy potential, do not require high altitudes. Accordingly, we limit this dataset to the first 6 staggered and 5 centred vertical layers, substantially reducing the file size. The data with all layers are available upon request.

\begin{figure}[H]
\centering
\includegraphics[width=0.9\textwidth]{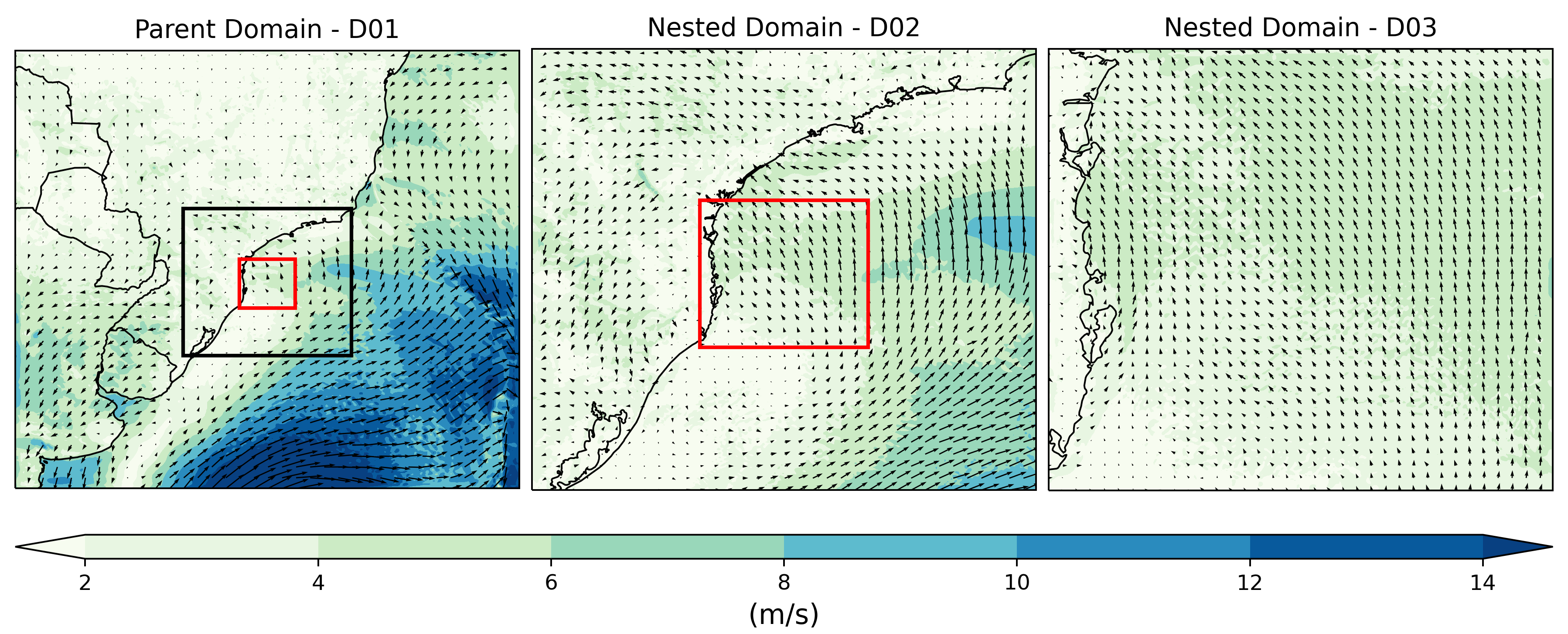}
\caption{Maps of the wind fields at 10 m from the 9km (D01), 3km (D02) and 1 km (D03) grids of the August 1\textsuperscript{st}, 2017 run of the WRF model.}\label{fig_wind10m}
\end{figure}

\begin{figure}[H]
\centering
\includegraphics[width=0.9\textwidth]{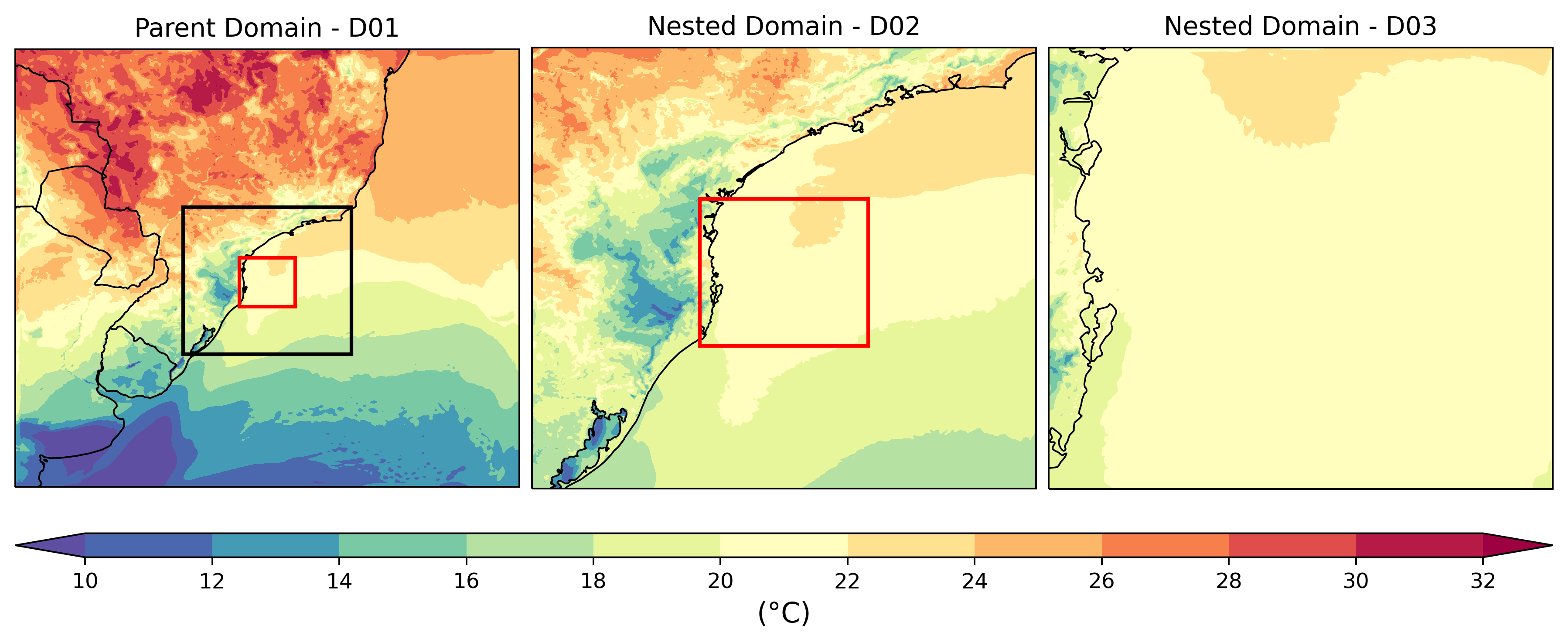}
\caption{Maps of 2-metre temperature fields from the 9 km (D01), 3 km (D02) and 1 km (D03) grids of the August 1\textsuperscript{st}, 2017, run of the WRF model.}\label{fig_temp2m}
\end{figure}

As an illustrative case, Figures \ref{fig_wind10m} and \ref{fig_temp2m} present the near-surface meteorological fields simulated across the three model domains for the experiment initialized on August 1\textsuperscript{st}, 2017. These figures specifically depict the spatial distribution of 2-metre air temperature and 10-metre wind.

\paragraph{SAR/CMOD5 data}

\begin{figure}[H]
\centering
\includegraphics[width=0.9\textwidth]{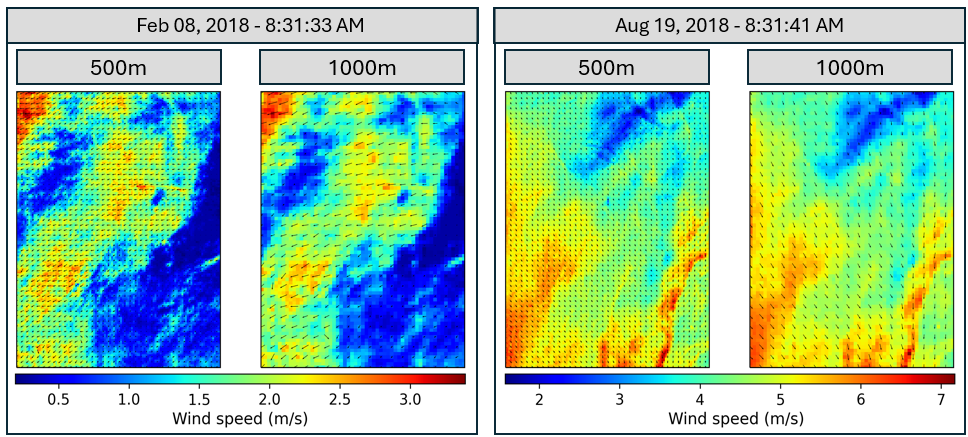}
\caption{SAR/CMOD5 Wind speed field maps derived from the SAR images of February 8, 2018, and August 19, 2018, at 500 m and 1 km spatial resolution.}\label{fig:sar_example}
\end{figure}

All the processed SAR/CMOD5 data are consolidated into a single NetCDF file containing 104 acquisition dates between 07 February 2017 and 11 November 2018. SAR-derived wind fields computed via CMOD5 inversion after masking and inpainting. The variables present in the SAR/CMOD5 data are shown in Table \ref{tab:sar_variables}. Two data files were created on the basis of spatial resolution (Table \ref{tab:sar_files_summary}). Examples of the wind speed fields derived from the model at both resolutions are shown in Figure \ref{fig:sar_example}.

\begin{table}[ht]
\small
  \centering
  \caption{Description of SAR/CMOD5 variables available in the dataset.}
  \label{tab:sar_variables}
  \begin{tabular}{lll}
    \hline
    \textbf{Variable} & \textbf{Description and units} & \textbf{Dimensions} \\
    \hline
    Times    & Date of SAR image acquisition (calendar date) & 1D (Time) \\
    XLAT     & Latitude values (degree\_north) & 2D (south\_north, west\_east) \\
    XLONG    & Longitude values (degree\_east) & 2D (south\_north, west\_east) \\
    XSAT     & Satellite identifier & 1D (satellite) \\
    U10      & 10 m U-component of wind ($\mathrm{m}\!\cdot\!\mathrm{s}^{-1}$) & 3D (Time, south\_north, west\_east) \\
    V10      & 10 m V-component of wind ($\mathrm{m}\!\cdot\!\mathrm{s}^{-1}$) & 3D (Time, south\_north, west\_east) \\
    WIND\_SPD & Wind speed magnitude ($\mathrm{m}\!\cdot\!\mathrm{s}^{-1}$) & 3D (Time, south\_north, west\_east) \\
    WIND\_DIR & Wind direction (degrees from north) & 3D (Time, south\_north, west\_east) \\
    \hline
  \end{tabular}
\end{table}

\begin{table}[ht]
  \centering
  \caption{Summary of SAR/CMOD5 NetCDF files.}
  \label{tab:sar_files_summary}
  \begin{tabular}{c c l}
    \hline
    \textbf{File Name} &  \textbf{Size} & \textbf{Description} \\
    \hline
    \texttt{data\_SAR\_500m.nc} & 7.47 MB
      & SAR/CMOD5 wind speed fields at 500 m spatial resolution. \\
    \texttt{data\_SAR\_1000m.nc} & 29.83 MB
      & SAR/CMOD5 wind speed fields at 1-km spatial resolution. \\
    \hline
  \end{tabular}
\end{table}

\paragraph{Itaja{\'i} oceanographic buoy}

In situ observations from the Itaja{\'i} metoceanographic buoy (PNBOIA program) are also used to evaluate the consistency of these results. The observations include date, time, geographic coordinates, wind speed, and wind direction. However, the observations are at a height of approximately 4 m. To compare them to the 10-m wind fields available in the WRF and SAR/CMOD5 outputs, we used the logarithmic profile law (Equation \ref{eq:log_prof}) \citep{emeis2018wind}:

\begin{equation}
  u(z)
  \;=\;
  u\bigl(z_{\mathrm{ref}}\bigr)\,
  \frac{\ln\!\bigl(\tfrac{z}{z_{0}}\bigr)}
       {\ln\!\bigl(\tfrac{z_{\mathrm{ref}}}{z_{0}}\bigr)}.
       \label{eq:log_prof}
\end{equation}

where $z$ is the target height for wind speed estimation [m], $z_{\mathrm{ref}}$ is the reference height at which the speed $u(z_{\mathrm{ref}})$ was measured [m], $u(z)$ is the wind speed at height $z$ [m/s], and $z_{0}$ is the surface roughness length [m]. The surface roughness length was estimated following the ERA5 methodology \citep{hersbach2023era5}. Given the relevance of the buoy data for the region-located within the WRF and SAR/CMOD5 study areas, these measurements are also provided in tabular format, with one row per observation time. The data is stored in the file \texttt{itajai\_buoy.csv} (2.17 MB), and the variables (columns) are listed in Table \ref{tab:buoy_variables}.

\begin{table}[ht]
\small
  \centering
  \caption{Description of buoy variables available in the dataset.}
  \label{tab:buoy_variables}
  \begin{tabular}{ll}
    \hline
    \textbf{Variable} & \textbf{Description and units} \\
    \hline
    Time      & Date and time of observation \\
    Latitude  & Latitude position of buoy (degree\_north) \\
    Longitude & Longitude position of buoy (degree\_east) \\
    Wdir      & Wind direction (degrees from north) \\
    Wspd      & Wind speed ($\mathrm{m}\!\cdot\!\mathrm{s}^{-1}$) \\
    Fsr       & Surface roughness length [m] \\
    Wspd\_10      & Wind speed adjusted to 10 m height ($\mathrm{m}\!\cdot\!\mathrm{s}^{-1}$) \\
    is\_SAR   & SAR comparison flag (Subsection \ref{subsec:tech_buoy}) \\
    \hline
  \end{tabular}
\end{table}

\section{Technical Validation}\label{sec:technical}

The data products presented in this dataset were evaluated using two approaches over the full observation period. First, we compared the 10-metre wind fields from the WRF model and the SAR/CMOD5 retrievals. Afterwards, we validated both datasets using in situ wind measurements from the Itaja{\'i} oceanographic buoy by comparing them with the nearest grid points in the WRF and SAR/CMOD5 data.

\subsection{WRF and SAR/CMOD5 grid comparison}

The wind speed data produced by the innermost WRF numerical model domain were compared with the results obtained by the SAR/CMOD5 model on the days/hours of image acquisition. Two different approaches were used for validation: a global analysis of the grid points and an analysis of the temporal mean. The 1-km SAR data and D03 WRF domain were used. Because both have the same spatial resolution and SAR/CMOD5 interpolated values to match the WRF grid, we have $82 \times 57$ values at each time (Table \ref{tab:grid_sizes}) and $104$ available SAR images. Therefore, there is a total of $N = 486,096$ records.


\paragraph{Global analysis.} For comparison, the WRF model outputs corresponding to the times closest to the SAR image acquisitions were used. The SAR images analysed in this study were acquired between 08:31:26 and 08:32:14 UTC (HH:MM:SS). Because the WRF model provides outputs at 30-minute intervals (i.e., at 00:00 and 30:00 UTC each hour), the 08:30:00 UTC output on the respective acquisition days was selected for comparison. Table~\ref{tab:wrf_sar_statistics} shows that the mean residual between the SAR/CMOD5 and WRF results (SAR/CMOD5 - WRF) is low (0.145 m/s), and the 5\textsuperscript{th} and 95\textsuperscript{th} percentiles indicate that 90\% of the residuals lie within the range \(\left[-3.297,\ 2.366\right]\), reflecting overall agreement between the datasets. However, the standard deviation of the residuals (1.782 m/s) suggests notable local discrepancies.

\begin{table}[ht]
  \centering
  \caption{Statistics of the WRF, SAR/CMOD5 data and their difference}
  \label{tab:wrf_sar_statistics}
  \begin{tabular}{l r r r}
    \hline
                        & \textbf{WRF} & \textbf{SAR/CMOD5} & \textbf{Residuals} \\
    \hline
    Minimum             &   0.080 &   0.200 &  -10.635 \\
    Maximum             &  16.806 &  19.775 &   11.541 \\
    Mean                &   6.719 &   6.506 &   -0.213 \\
    Standard deviation  &   2.813 &   2.687 &    1.782 \\
    1\textsuperscript{st} percentile  &   1.334 &   0.580 &   -5.022 \\
    5\textsuperscript{th} percentile  &   2.246 &   2.427 &   -3.297 \\
    95\textsuperscript{th} percentile &  11.380 &  11.349 &    2.366 \\
    99\textsuperscript{th} percentile &  14.324 &  14.104 &    3.647 \\
    \hline
  \end{tabular}
\end{table}

\begin{figure}[H]
\centering
\includegraphics[width=\textwidth]{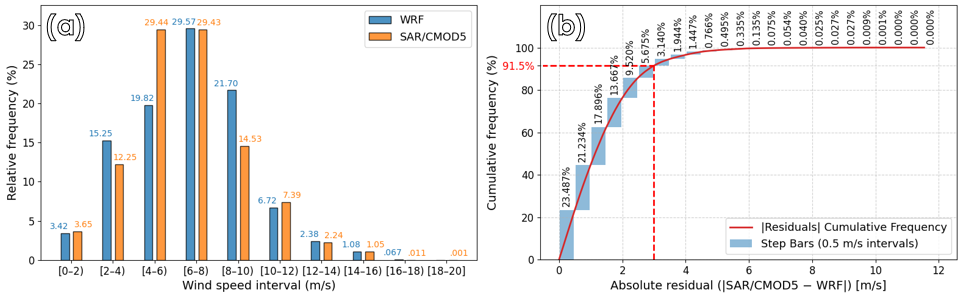}
\caption{(a) Comparison of 10-m wind speed data from WRF and SAR/CMOD5 at the times of SAR image acquisition. (a) Relative frequency distributions of average wind speeds for the WRF (blue) and SAR/CMOD5 (orange) data across 2 m/s intervals. (b) Cumulative distribution of absolute residuals, including a step-bar representation of the incremental contributions within 0.5 m/s bins and the cumulative frequency curve.}\label{fig:wrf_sar_comparison}
\end{figure}

The distributions are shown in Figure \ref{fig:wrf_sar_comparison}. The distributions of the wind intensity values in each pixel for the WRF and SAR/CMOD5 models are shown in Figure \ref{fig:wrf_sar_comparison}a. The distributions are shown in Figure~\ref{fig:wrf_sar_comparison}. The relative frequency distributions of the wind speed values (per pixel) for both the WRF and the SAR/CMOD5 data are shown in Figure~\ref{fig:wrf_sar_comparison}a. The distributions exhibit similar overall patterns, with the highest frequencies occurring in the [4-6) and [6-8) m/s intervals for both models. SAR/CMOD5 has a heavier right tail, whereas WRF has greater values in the lower range ([0-4) m/s), which is consistent with the results in Table~\ref{tab:wrf_sar_statistics}.

The cumulative distributions of the absolute residuals between the SAR/CMOD5 and WRF models are shown in Figure~\ref{fig:wrf_sar_comparison}b. More than 90\% of the absolute residuals are below 3 m/s, indicating good agreement between the distributions. This is supported by the error metrics \citep{murphy2012machine}, with an RMSE (Eq. \ref{eq:rmse}) of 1.795 m/s and an MAE (Eq. \ref{eq:mae}) of 1.388 m/s.

\begin{equation}
  \text{root mean squared error (}\mathrm{RMSE}\text{)}
  = \sqrt{\frac{1}{N} \sum_{t=1}^{N} residual(t)^2}
  \label{eq:rmse}
\end{equation}

\begin{equation}
  \text{mean absolute error (}\mathrm{MAE}\text{)}
  = \frac{1}{N} \sum_{t=1}^{N} \bigl|redisual(t)\bigr|
  \label{eq:mae}
\end{equation}

\begin{figure}[H]
\centering
\includegraphics[width=\textwidth]{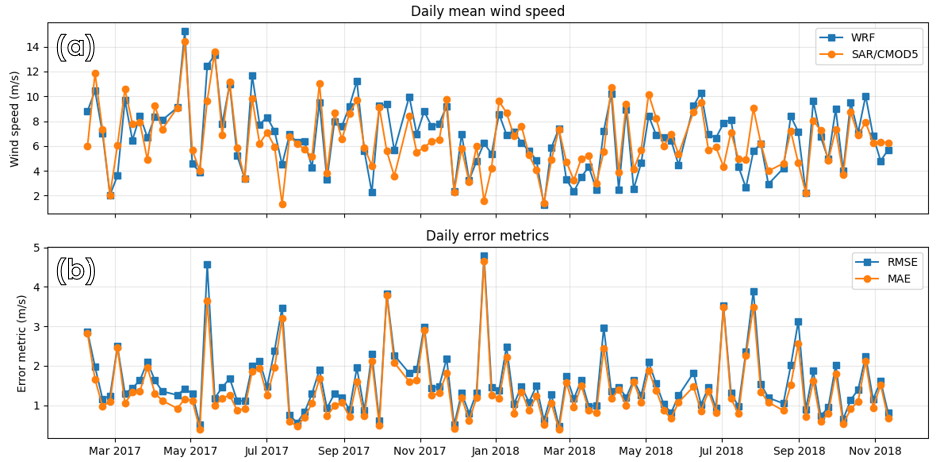}
\caption{Daily mean wind speed at 10 m (m/s) from the WRF and SAR/CMOD5 models over 104 SAR acquisitions spaced at 6-day intervals between February 2017 and November 2018. The upper panel shows the time series of daily spatial averages from both models. The lower panel presents the corresponding RMSE and MAE for each day.}
\label{fig:wrf_sar_daily_series}

\end{figure}

\paragraph{Temporal mean.} In the second approach, the WRF and SAR/CMOD5 wind speeds were compared on the basis of the daily mean across all pixels ($82 \times 57 = 4674$) within the region. The upper panel of Figure~\ref{fig:wrf_sar_daily_series} presents the daily mean wind speeds from both models over the 104 SAR acquisition days, while the lower panel displays the corresponding daily error metrics. As shown in the figure and summarized in Table~\ref{tab:rmse_mae_thresholds}, at the stricter 1 m/s threshold, over one-third of the days (35.6\%) have MAE values within the limit, whereas only 21.2\% meet this criterion for the RMSE. This indicates stronger local deviations that significantly affect the RMSE, which is more sensitive to outliers. Nonetheless, more than 93\% of the days exhibit RMSE and MAE values less than 3 m/s. Overall, the results indicate good agreement between the models in terms of daily spatial averages.

\begin{table}[ht]
  \centering
  \caption{Proportion of days used in the comparison with RMSE and MAE below given thresholds.}
  \label{tab:rmse_mae_thresholds}
  \begin{tabular}{c c c}
    \hline
    \textbf{Threshold (m/s)} & \textbf{RMSE $\le$ Threshold} & \textbf{MAE $\le$ Threshold} \\
    \hline
    1 & 22/104 (21.2\%)  & 37/104 (35.6\%)  \\
    2 & 81/104 (77.9\%)  & 88/104 (84.6\%)  \\
    3 & 97/104 (93.3\%)  & 98/104 (94.2\%)  \\
    5 & 104/104 (100.0\%) & 104/104 (100.0\%) \\
    \hline
  \end{tabular}
\end{table}

\subsection{Buoy comparison}\label{subsec:tech_buoy}

To evaluate the accuracy of the WRF model and SAR/CMOD5 wind speed estimates, we compare them with measurements from the Itaja{\'i} buoy. The buoy coordinates vary slightly over time, but for consistency, we used a fixed position of 27$\degree$24.35'S, 47$\degree$15.93'W. The wind speed values from the buoy were adjusted to represent the 10-metre height.

\paragraph{WRF}

The WRF simulation data spans from 2017-02-06 06:30 UTC to 2018-11-18 06:00 UTC. Buoy observations are recorded hourly at hh:00:00 UTC. Within the entire WRF simulation period, there are 59 missing records in the buoy dataset: one isolated missing hour on 2018-04-15 at 14:00 UTC and one continuous gap of 58 hours from 2018-04-30 at 01:00 UTC to 2018-05-02 at 10:00 UTC (inclusive). After accounting for these gaps, there are 15,497 timestamps with overlapping WRF and buoy wind speed data. Because the closest grid point
in the WRF data was used for comparison, 15,497 is also the number of observations used in the comparison.

\begin{figure}[H]
\centering
\includegraphics[width=\textwidth]{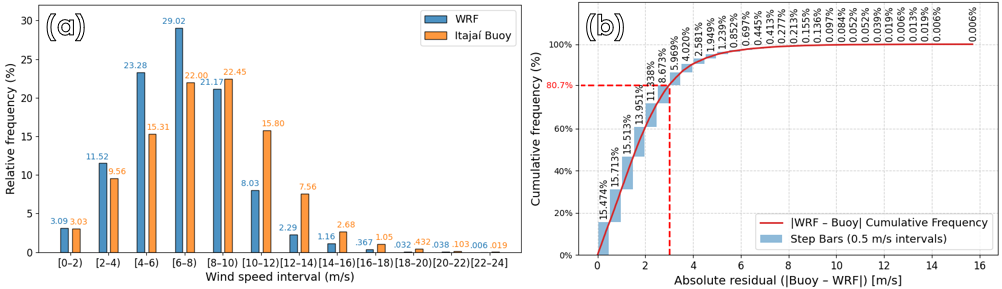}
\caption{(a) Comparison of 10 m wind speed data from WRF and SAR/CMOD5 at the times of SAR image acquisition. (a) Relative frequency distributions of average wind speeds for WRF (blue) and Itaja{\'i} buoy (orange) data across 2 m/s intervals. (b) Cumulative distribution of absolute residuals, including a step-bar representation of the incremental contributions within 0.5 m/s bins and the cumulative frequency curve.} \label{fig:buoy_wrf_comparison}
\end{figure}

The wind speed distributions from the WRF model and the Itaja{\'i} buoy are compared in Figure~\ref{fig:buoy_wrf_comparison}. Panel (a) shows the relative frequency of the wind speed values binned at 2 m/s intervals, revealing distinct distributions--most notably, a heavier right tail in the buoy data. Despite these differences, the cumulative distribution of absolute residuals in panel (b) indicates that more than 80\% of the absolute residuals fall below 3 m/s. The calculated RMSE and MAE are 2.538 m/s and 1.956 m/s, respectively.

\paragraph{SAR/CMOD5}

\begin{figure}[H]
\centering
\includegraphics[width=0.9\textwidth]{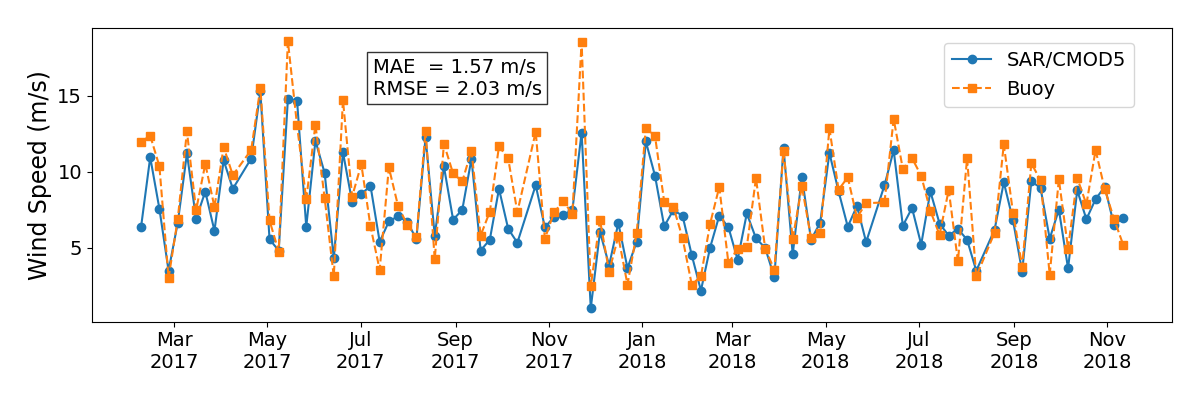}
\caption{Time series of wind speed (m/s) from the SAR/CMOD5 closest grid point to the Itaja{\'i} buoy and buoy measurements adjusted to a height of 10 metres. The plot includes 104 SAR acquisitions at 6-day intervals between February 2017 and November 2018.}\label{fig:buoy_sar}
\end{figure}

The buoy provides hourly observations (hh:00:00), and because the SAR acquisitions during the study period occurred between 08:31:26 and 08:32:14 UTC, the closest hourly record at 09:00 UTC was used for comparison. Similarly, the nearest grid point in the SAR data was selected. The good agreement between the SAR-derived wind speeds and the buoy observations across the entire time series of 104 SAR acquisitions spaced at 6-day intervals is shown in Figure \ref{fig:buoy_sar}. An MAE of 1.57 m/s and a RMSE of 2.03 m/s support these results, with a higher RMSE reflecting the influence of larger individual deviations between the SAR and buoy measurements.

\newpage
\section{Usage notes}

The dataset files are provided in NetCDF (\texttt{.nc}) format, a widely used standard for array-oriented scientific data. This format allows for easy manipulation and analysis using libraries such as \texttt{xarray} in Python. To work with the dataset, you can load a file named "\texttt{filename.nc}" using the following code snippet:

\begin{python}
import xarray as xr
dataset = xr.open_dataset("data_D01_20170206.nc")
u_wind_component = dataset["U"]
\end{python}

The code behind Section \ref{sec:technical}, along with example scripts demonstrating key workflows (e.g., loading WRF output, computing heights, and generating figures), is available on GitHub (https://github.com/luancvieira/Southwestern-South-Atlantic-weather-dataset).

\section*{Data availability}

The dataset is accessible in the Harvard Dataverse repository: \url{https://dataverse.harvard.edu/dataset.xhtml?persistentId=doi:10.7910/DVN/OUEZ0H}. The data with all 45 vertical layers is available upon request.

\section*{Code availability}

The dataset was produced using the Weather Research and Forecasting (WRF) model, which is accessible via its Git repository (\url{https://github.com/wrf-model/WRF}). Further information can be obtained from the official website of the model (\url{https://www2.mmm.ucar.edu/wrf/users}). SAR data processing relied on the XSAR (\url{https://github.com/umr-lops/xsar/tree/develop}) and XSARSEA (\url{https://github.com/umr-lops/xsarsea}) Python libraries, which offer tools for reading, processing, and analyzing Sentinel-1 SAR measurements for ocean and coastal applications.

\bibliography{references}

@inproceedings{bertalmio2001navier,
  title={Navier-Stokes, fluid dynamics, and image and video inpainting},
  author={Bertalmio, Marcelo and Bertozzi, Andrea L. and Sapiro, Guillermo},
  booktitle={Proceedings of the 2001 IEEE Computer Society Conference on Computer Vision and Pattern Recognition (CVPR)},
  volume={1},
  pages={I--I},
  year={2001},
  organization={IEEE}
}

@article{telea2004image,
  title={An image inpainting technique based on the fast marching method},
  author={Telea, Alexandru},
  journal={Journal of Graphics Tools},
  volume={9},
  number={1},
  pages={23--34},
  year={2004},
  publisher={Taylor \& Francis}
}

@techreport{hersbach2003cmod5,
  author={Hersbach, Hans},
  title={CMOD5-An improved geophysical model function for ERS C-band scatterometry},
  institution={ECMWF},
  year={2003},
  number={Tech. Memo. 395}
}

@inproceedings{wackerman2003automated,
  title={Automated estimation of wind vectors from SAR},
  author={Wackerman, C. C. and Pichel, W. G. and Clemente-Colon, P.},
  booktitle={Proceedings of the 12th Conference on Interactions of the Sea and Atmosphere},
  year={2003}
}

@techreport{skamarock_etal_2021,
  author={Skamarock, W. C. et al.},
  city={Boulder},
  doi={10.5065/1dfh-6p97},
  institution={NCAR},
  journal={NCAR Technical Notes},
  title={A Description of the Advanced Research WRF Model Version 4.3},
  url={https://opensky.ucar.edu/islandora/object/opensky:2898},
  year={2021}
}

@techreport{Mesinger_Arakawa_1976,
  author={Mesinger, F. and Arakawa, A.},
  title={Numerical Methods Used in Atmospheric Model, Volume 1},
  institution={International Council of Scientific Unions-World Meteorological Organization},
  year={1976}
}

@article{Lundquist_etal_2010,
  author={Katherine A. Lundquist and Fotini Katopodes Chow and Julie K. Lundquist},
  title={An Immersed Boundary Method for the Weather Research and Forecasting Model},
  journal={Monthly Weather Review},
  year={2010},
  publisher={American Meteorological Society},
  address={Boston, MA, USA},
  volume={138},
  number={3},
  doi={https://doi.org/10.1175/2009MWR2990.1},
  pages={796--817},
  url={https://journals.ametsoc.org/view/journals/mwre/138/3/2009mwr2990.1.xml}
}

@article { morrison_2009,
      author = "H. Morrison and G. Thompson and V. Tatarskii",
      title = "Impact of Cloud Microphysics on the Development of Trailing Stratiform Precipitation in a Simulated Squall Line: Comparison of One- and Two-Moment Schemes",
      journal = "Monthly Weather Review",
      year = "2009",
      publisher = "American Meteorological Society",
      address = "Boston MA, USA",
      volume = "137",
      number = "3",
      doi = "https://doi.org/10.1175/2008MWR2556.1",
      pages=      "991 - 1007",
      url = "https://journals.ametsoc.org/view/journals/mwre/137/3/2008mwr2556.1.xml"
}

@phdthesis{dragaud_2021,
  author={Dragaud, I. C. D. V.},
  title={An{\'a}lise dos mecanismos f{\'i}sicos que governam o escoamento atmosf{\'e}rico na regi{\~a}o costeira do estado do Rio de Janeiro},
  school={Universidade Federal do Rio de Janeiro},
  year={2021}
}

@article{soares_da_silva_2023,
  author={Soares da Silva, M. and Pimentel, Luiz Claudio Gomes and Duda, Fernando Pereira and Arag{\~a}o, Leonardo and Silva, Corbiniano and Dragaud, Ian Cunha D’Amato Viana and Vicentini, Pedro Caffaro},
  title={Assessment of meteorological settings on air quality modeling system-a proposal for UN-SDG and regulatory studies in non-homogeneous regions in Brazil},
  journal={Environmental Science and Pollution Research},
  volume={30},
  pages={1737--1760},
  year={2023},
  doi={10.1007/s11356-022-22146-1},
  url={https://doi.org/10.1007/s11356-022-22146-1}
}

@article{deSouzaetal_2024,
  author={de Souza, Lucio Silva and Soares da Silva, Mauricio and de Almeida, Vinicius Albuquerque and Moraes, Nilton Oliveira and de Souza, Enio Pereira and Senna, M{\^o}nica Carneiro Alves and Fran{\c c}a, Gutemberg Borges and Frota, Maur{\'i}cio Nogueira and de Almeida, Manoel Valdonel and Viana, Lude Quieto},
  title={Evaluation of Cumulus and Microphysical Parameterization Schemes of the WRF Model for Precipitation Prediction in the Para{\'\i}ba do Sul River Basin, Southeastern Brazil},
  journal={Pure and Applied Geophysics},
  volume={181},
  pages={679--700},
  year={2024},
  doi={10.1007/s00024-023-03419-3},
  url={https://doi.org/10.1007/s00024-023-03419-3}
}

@misc{usgs_2023,
  author={USGS - United States Geological Survey},
  title={USGS EROS Archive-Digital Elevation-Shuttle Radar Topography Mission (SRTM) 1 Arc-Second Global},
  note={Accessed on October 2023},
  url={https://www.usgs.gov/centers/eros/science/usgs-eros-archive-digital-elevation-shuttle-radar-topography-mission-srtm-1?qt-science_center_objects=0#qt-science_center_objects},
  year={2023}
}

@Article{MapBiomas,
  author={Souza, Carlos M. and Z. Shimbo, Julia and Rosa, Marcos R. and Parente, Leandro L. and A. Alencar, Ane and Rudorff, Bernardo F. T. and Hasenack, Heinrich and Matsumoto, Marcelo and G. Ferreira, Laerte and Souza-Filho, Pedro W. M. and de Oliveira, Sergio W. and Rocha, Washington F. and Fonseca, Ant{\'o}nio V. and Marques, Camila B. and Diniz, Cesar G. and Costa, Diego and Monteiro, Dyeden and Rosa, Eduardo R. and V{\'e}lez-Martin, Eduardo and Weber, Eliseu J. and Lenti, Felipe E. B. and Paternost, Fernando F. and Pareyn, Frans G. C. and Siqueira, Jo{\~a}o V. and Viera, Jos{\'e} L. and Neto, Luiz C. Ferreira and Saraiva, Marciano M. and Sales, Marcio H. and Salgado, Moises P. G. and Vasconcelos, Rodrigo and Galano, Soltan and Mesquita, Vinicius V. and Azevedo, Tasso},
  title={Reconstructing Three Decades of Land Use and Land Cover Changes in Brazilian Biomes with Landsat Archive and Earth Engine},
  journal={Remote Sensing},
  volume={12},
  year={2020},
  number={17},
  article-number={2735},
  url={https://www.mdpi.com/2072-4292/12/17/2735},
  issn={2072-4292},
  doi={10.3390/rs12172735}
}

@article{Pedruzzi_2022,
  author={Rizzieri Pedruzzi and Willian Lemker Andre{\~a}o and Bok Haeng Baek and Anderson Paulo Hudke and Timothy William Glotfelty and Edmilson Dias de Freitas and Jorge Alberto Martins and Jared H. Bowden and Janaina Antonino Pinto and Marcelo Felix Alonso and Taciana Toledo de Almeida Abuquerque},
  title={Update of land use/land cover and soil texture for Brazil: Impact on WRF modeling results over Para{\'\i}ba do Sul River Basin, S{\~a}o Paulo},
  journal={Atmospheric Environment},
  volume={268},
  pages={118760},
  year={2022},
  doi={https://doi.org/10.1016/j.atmosenv.2021.118760},
  url={https://www.sciencedirect.com/science/article/pii/S1352231021005823}
}

@mastersthesis{Jacinto_2020,
  author={Jacinto, Larissa de Freitas Ramos},
  title={Sistema Integrado de Modelagem Ambiental dos Fen{\^o}menos de Transporte na {\'A}rea de Influ{\^e}ncia de Centrais Nucleares},
  school={Programa de P{\'o}s-Gradua{\c c}{\~a}o em Meteorologia, PPGM, Universidade Federal do Rio de Janeiro},
  year={2020}
}

@TechReport{Danielson2011,
  author={Danielson, Jeffrey J. and Gesch, Dean B.},
  editor={U.S. Geological Survey},
  title={Global multi-resolution terrain elevation data 2010 (GMTED2010)},
  series={Open-File Report},
  year={2011},
  edition={--},
  note={Report},
  issn={2011-1073},
  doi={10.3133/ofr20111073},
  url={https://doi.org/10.3133/ofr20111073},
  language={ENGLISH}
}

@misc{MCD12Q2_006,
  author={Mark A. Friedl and Josh Gray and Damien Sulla-Menashe},
  title={MCD12Q2 MODIS/Terra+Aqua Land Cover Dynamics Yearly L3 Global 500m SIN Grid Version 6},
  year={2019},
  publisher={NASA EOSDIS Land Processes DAAC},
  doi={10.5067/MODIS/MCD12Q2.006},
  url={https://doi.org/10.5067/MODIS/MCD12Q2.006},
  note={Accessed on 2025-06-04}
}

@ARTICLE{lellouche_etal2021,
  author={Jean-Michel, Lellouche and Eric, Greiner and Romain, Bourdall{\'e}-Badie and Gilles, Garric and Ang{\'e}lique, Melet and Marie, Dr{\'e}villon and Cl{\'e}ment, Bricaud and Mathieu, Hamon and Olivier, Le Galloudec and Charly, Regnier and Tony, Candela and Charles-Emmanuel, Testut and Florent, Gasparin and Giovanni, Ruggiero and Mounir, Benkiran and Yann, Drillet and Pierre-Yves, Le Traon},
  title={The Copernicus Global 1/12° Oceanic and Sea Ice GLORYS12 Reanalysis},
  journal={Frontiers in Earth Science},
  volume={9},
  year={2021},
  doi={10.3389/feart.2021.698876},
  issn={2296-6463}
}

@Article{macferrin_etal2025,
  author={MacFerrin, M. and Amante, C. and Carignan, K. and Love, M. and Lim, E.},
  title={The Earth Topography 2022 (ETOPO 2022) global DEM dataset},
  journal={Earth System Science Data},
  volume={17},
  number={5},
  pages={1835--1849},
  year={2025},
  url={https://essd.copernicus.org/articles/17/1835/2025/},
  doi={10.5194/essd-17-1835-2025}
}

@Article{good_etal2020,
  author={Good, Simon and Fiedler, Emma and Mao, Chongyuan and Martin, Matthew J. and Maycock, Adam and Reid, Rebecca and Roberts-Jones, Jonah and Searle, Toby and Waters, Jennifer and While, James and Worsfold, Mark},
  title={The Current Configuration of the OSTIA System for Operational Production of Foundation Sea Surface Temperature and Ice Concentration Analyses},
  journal={Remote Sensing},
  volume={12},
  number={4},
  article-number={720},
  year={2020},
  issn={2072-4292},
  doi={10.3390/rs12040720}
}

@Article{egbert_erofeeva2002,
  author={Gary D. Egbert and Svetlana Y. Erofeeva},
  title={Efficient Inverse Modeling of Barotropic Ocean Tides},
  journal={Journal of Atmospheric and Oceanic Technology},
  volume={19},
  number={2},
  pages={183--204},
  year={2002}
}

@Article{egbert_etal1994,
  author={Egbert, Gary D. and Bennett, Andrew F. and Foreman, Michael G. G.},
  title={TOPEX/POSEIDON tides estimated using a global inverse model},
  journal={Journal of Geophysical Research: Oceans},
  volume={99},
  number={C12},
  pages={24821--24852},
  year={1994},
  issn={2156-2202},
  doi={10.1029/94JC01894}
}

@article{shchepetkin_mcwilliams2005,
  author={Shchepetkin, A. F. and McWilliams, J. C.},
  journal={Ocean Modelling},
  pages={347--404},
  title={The regional oceanic modeling system (ROMS): a split-explicit, free-surface, topography-following-coordinate oceanic model},
  volume={9},
  year={2005}
}

@misc{hersbach2023era5,
  title={ERA5 hourly data on single levels from 1940 to present, Copernicus Climate Change Service (C3S) Climate Data Store (CDS) [data set]},
  author={Hersbach, Hans and Bell, Bill and Berrisford, Paul and Biavati, Gionata and Hor{\'a}nyi, Andr{\'a}s and Mu{\~n}oz Sabater, Joaqu{\'\i}n and Nicolas, Julien and Peubey, Carole and Radu, Raluca and Rozum, Iryna and others},
  year={2023}
}

@article{chidichimo2023energetic,
  title={Energetic overturning flows, dynamic interocean exchanges, and ocean warming observed in the South Atlantic},
  author={Chidichimo, Mar{\'\i}a Paz and Perez, Renellys C. and Speich, Sabrina and Kersal{\'e}, Marion and Sprintall, Janet and Dong, Shenfu and Lamont, Tarron and Sato, Olga T. and Chereskin, Teresa K. and Hummels, Rebecca and others},
  journal={Communications Earth \& Environment},
  volume={4},
  number={1},
  pages={10},
  year={2023},
  publisher={Nature Publishing Group UK London}
}

@article{gonzalez2020regulation,
  title={Regulation for offshore wind power development in Brazil},
  author={Gonz{\'a}lez, Mario Orestes Aguirre and Santiso, Andressa Medeiros and de Melo, David Cassimiro and de Vasconcelos, Rafael Monteiro},
  journal={Energy Policy},
  volume={145},
  pages={111756},
  year={2020},
  publisher={Elsevier}
}

@article{vinhoza2021brazil,
  title={Brazil's offshore wind energy potential assessment based on a Spatial Multi-Criteria Decision Analysis},
  author={Vinhoza, Amanda and Schaeffer, Roberto},
  journal={Renewable and Sustainable Energy Reviews},
  volume={146},
  pages={111185},
  year={2021},
  publisher={Elsevier}
}

@article{azevedo2020assessment,
  title={Assessment of offshore wind power Potential along the Brazilian coast},
  author={Azevedo, Sylvester Stallone Pereira de and Junior, Amaro Olimpio Pereira and Silva, Neilton Fidelis da and Ara{\'u}jo, Renato Samuel Barbosa de and J{\'u}nior, Antonio Ald{\'\i}sio Carlos},
  journal={Energies},
  volume={13},
  number={10},
  pages={2557},
  year={2020},
  publisher={MDPI}
}

@article{catolico2021socioeconomic,
  title={Socioeconomic impacts of large hydroelectric power plants in Brazil: A synthetic control assessment of Estreito hydropower plant},
  author={Catolico, A. C. C. and Maestrini, M. and Strauch, J. C. M. and Giusti, F. and Hunt, J.},
  journal={Renewable and Sustainable Energy Reviews},
  volume={151},
  pages={111508},
  year={2021},
  publisher={Elsevier}
}

@article{bastos2023wave,
  title={Wave energy generation in Brazil: A georeferenced oscillating water column inventory},
  author={Bastos, Adriano Silva and Souza, T{\^a}mara Rita Costa de and Ribeiro, Dieimys Santos and Melo, Mirian de Lourdes Noronha Motta and Martinez, Carlos Barreira},
  journal={Energies},
  volume={16},
  number={8},
  pages={3409},
  year={2023},
  publisher={MDPI}
}

@ARTICLE{6504845,
  author={Moreira, Alberto and Prats-Iraola, Pau and Younis, Marwan and Krieger, Gerhard and Hajnsek, Irena and Papathanassiou, Konstantinos P.},
  title={A tutorial on synthetic aperture radar},
  journal={IEEE Geoscience and Remote Sensing Magazine},
  year={2013},
  volume={1},
  number={1},
  pages={6--43},
  doi={10.1109/MGRS.2013.2248301}
}

@article{ryu2025increasing,
  title={Increasing resolution and accuracy in sub-seasonal forecasting through 3D U-Net: the Western US},
  author={Ryu, Jihun and Kim, Hisu and Wang, Shih-Yu Simon and Yoon, Jin-Ho},
  journal={EGUsphere},
  volume={2025},
  pages={1--18},
  year={2025},
  publisher={Copernicus Publications G{\"o}ttingen, Germany}
}

@article{hadjipetrou2024high,
  title={High-Resolution Wind Speed Estimates for the Eastern Mediterranean Basin: A Statistical Comparison Against Coastal Meteorological Observations},
  author={Hadjipetrou, Stylianos and Kyriakidis, Phaedon},
  journal={Wind},
  volume={4},
  number={4},
  pages={311--341},
  year={2024},
  publisher={MDPI}
}

@article{hadjipetrou2022evaluating,
  title={Evaluating the suitability of Sentinel-1 SAR data for offshore wind resource assessment around Cyprus},
  author={Hadjipetrou, Stylianos and Liodakis, Stelios and Sykioti, Anastasia and Katikas, Loukas and Park, No-Wook and Kalogirou, Soteris and Akylas, Evangelos and Kyriakidis, Phaedon},
  journal={Renewable Energy},
  volume={182},
  pages={1228--1239},
  year={2022},
  publisher={Elsevier}
}

@article{gramcianinov2023recent,
  title={Recent changes in extreme wave events in the south-western South Atlantic},
  author={Gramcianinov, Carolina B. and Staneva, Joanna and Souza, Celia R. G. and Linhares, Priscila and de Camargo, Ricardo and da Silva Dias, Pedro L.},
  journal={State of the Planet},
  volume={1},
  pages={1--12},
  year={2023},
  publisher={Copernicus GmbH}
}

@article{pagano2024challenges,
  title={Challenges of operational weather forecast verification and evaluation},
  author={Pagano, Thomas C. and Casati, Barbara and Landman, Stephanie and Loveday, Nicholas and Taggart, Robert and Ebert, Elizabeth E. and Khanarmuei, Mohammadreza and Jensen, Tara L. and Mittermaier, Marion and Roberts, Helen and others},
  journal={Bulletin of the American Meteorological Society},
  volume={105},
  number={4},
  pages={E789--E802},
  year={2024},
  publisher={American Meteorological Society}
}

@book{emeis2018wind,
  title={Wind energy meteorology: atmospheric physics for wind power generation},
  author={Emeis, Stefan},
  year={2018},
  publisher={Springer}
}

@book{murphy2012machine,
  title={Machine learning: a probabilistic perspective},
  author={Murphy, Kevin P.},
  year={2012},
  publisher={MIT Press}
}

@article{bracco2025machine,
  title={Machine learning for the physics of climate},
  author={Bracco, Annalisa and Brajard, Julien and Dijkstra, Henk A. and Hassanzadeh, Pedram and Lessig, Christian and Monteleoni, Claire},
  journal={Nature Reviews Physics},
  volume={7},
  number={1},
  pages={6--20},
  year={2025},
  publisher={Nature Publishing Group UK London}
}

@article{chen2023machine,
  title={Machine learning methods in weather and climate applications: A survey},
  author={Chen, Liuyi and Han, Bocheng and Wang, Xuesong and Zhao, Jiazhen and Yang, Wenke and Yang, Zhengyi},
  journal={Applied Sciences},
  volume={13},
  number={21},
  pages={12019},
  year={2023},
  publisher={MDPI}
}

@article{li2024generative,
  title={Generative emulation of weather forecast ensembles with diffusion models},
  author={Li, Lizao and Carver, Robert and Lopez-Gomez, Ignacio and Sha, Fei and Anderson, John},
  journal={Science Advances},
  volume={10},
  number={13},
  pages={eadk4489},
  year={2024},
  publisher={American Association for the Advancement of Science}
}

@article{wang2019fast,
  title={Fast domain-aware neural network emulation of a planetary boundary layer parameterization in a numerical weather forecast model},
  author={Wang, Jiali and Balaprakash, Prasanna and Kotamarthi, Rao},
  journal={Geoscientific Model Development},
  volume={12},
  number={10},
  pages={4261--4274},
  year={2019},
  publisher={Copernicus Publications G{\"o}ttingen, Germany}
}

@article{zhang2024deep,
  title={Deep-learning correction methods for weather research and forecasting (WRF) model precipitation forecasting: a case study over Zhengzhou, China},
  author={Zhang, Jianbin and Gao, Zhiqiu and Li, Yubin},
  journal={Atmosphere},
  volume={15},
  number={6},
  pages={631},
  year={2024},
  publisher={MDPI}
}

@article{zhu2023weather2k,
  title={Weather2k: A multivariate spatio-temporal benchmark dataset for meteorological forecasting based on real-time observation data from ground weather stations},
  author={Zhu, Xun and Xiong, Yutong and Wu, Ming and Nie, Gaozhen and Zhang, Bin and Yang, Ziheng},
  journal={arXiv preprint arXiv:2302.10493},
  year={2023}
}

@article{wang2022deep,
  title={On deep learning-based bias correction and downscaling of multiple climate models simulations},
  author={Wang, Fang and Tian, Di},
  journal={Climate Dynamics},
  volume={59},
  number={11},
  pages={3451--3468},
  year={2022},
  publisher={Springer}
}

@article{oh2022machine,
  title={Machine learning-based statistical downscaling of wind resource maps using multi-resolution topographical data},
  author={Oh, Myeongchan and Lee, Jehyun and Kim, Jin-Young and Kim, Hyun-Goo},
  journal={Wind Energy},
  volume={25},
  number={6},
  pages={1121--1141},
  year={2022},
  publisher={Wiley Online Library}
}

@article{wu2024data,
  title={Data-driven weather forecasting and climate modeling from the perspective of development},
  author={Wu, Yuting and Xue, Wei},
  journal={Atmosphere},
  volume={15},
  number={6},
  pages={689},
  year={2024},
  publisher={MDPI}
}

@article { Zheng_2016,
      author = "Yue Zheng and Kiran Alapaty and Jerold A. Herwehe and Anthony D. Del Genio and Dev Niyogi",
      title = "Improving High-Resolution Weather Forecasts Using the Weather Research and Forecasting (WRF) Model with an Updated Kain-Fritsch Scheme",
      journal = "Monthly Weather Review",
      year = "2016",
      publisher = "American Meteorological Society",
      address = "Boston MA, USA",
      volume = "144",
      number = "3",
      doi = "https://doi.org/10.1175/MWR-D-15-0005.1",
      pages=      "833 - 860",
      url = "https://journals.ametsoc.org/view/journals/mwre/144/3/mwr-d-15-0005.1.xml"
}

@article{iacono_2008,
author = {Iacono, Michael J. and Delamere, Jennifer S. and Mlawer, Eli J. and Shephard, Mark W. and Clough, Shepard A. and Collins, William D.},
title = {Radiative forcing by long-lived greenhouse gases: Calculations with the AER radiative transfer models},
journal = {Journal of Geophysical Research: Atmospheres},
volume = {113},
number = {D13},
pages = {},
doi = {https://doi.org/10.1029/2008JD009944},
url = {https://agupubs.onlinelibrary.wiley.com/doi/abs/10.1029/2008JD009944},
eprint = {https://agupubs.onlinelibrary.wiley.com/doi/pdf/10.1029/2008JD009944},
year = {2008}
}

@article { jimenez_2012,
      author = "Pedro A. Jim{\'e}nez and Jimy Dudhia and J. Fidel González-Rouco and Jorge Navarro and Juan P. Montávez and Elena García-Bustamante",
      title = "A Revised Scheme for the WRF Surface Layer Formulation",
      journal = "Monthly Weather Review",
      year = "2012",
      publisher = "American Meteorological Society",
      address = "Boston MA, USA",
      volume = "140",
      number = "3",
      doi = "https://doi.org/10.1175/MWR-D-11-00056.1",
      pages=      "898 - 918",
      url = "https://journals.ametsoc.org/view/journals/mwre/140/3/mwr-d-11-00056.1.xml"
}

@article { hong_2006,
      author = "Song-You Hong and Yign Noh and Jimy Dudhia",
      title = "A New Vertical Diffusion Package with an Explicit Treatment of Entrainment Processes",
      journal = "Monthly Weather Review",
      year = "2006",
      publisher = "American Meteorological Society",
      address = "Boston MA, USA",
      volume = "134",
      number = "9",
      doi = "https://doi.org/10.1175/MWR3199.1",
      pages=      "2318 - 2341",
      url = "https://journals.ametsoc.org/view/journals/mwre/134/9/mwr3199.1.xml"
}

@article{niu_2011,
author = {Niu, Guo-Yue and Yang, Zong-Liang and Mitchell, Kenneth E. and Chen, Fei and Ek, Michael B. and Barlage, Michael and Kumar, Anil and Manning, Kevin and Niyogi, Dev and Rosero, Enrique and Tewari, Mukul and Xia, Youlong},
title = {The community Noah land surface model with multiparameterization options (Noah-MP): 1. Model description and evaluation with local-scale measurements},
journal = {Journal of Geophysical Research: Atmospheres},
volume = {116},
number = {D12},
pages = {},
doi = {https://doi.org/10.1029/2010JD015139},
url = {https://agupubs.onlinelibrary.wiley.com/doi/abs/10.1029/2010JD015139},
eprint = {https://agupubs.onlinelibrary.wiley.com/doi/pdf/10.1029/2010JD015139},
year = {2011}
}

@article{bilgili2023potential,
  title={Potential visibility, growth, and technological innovation in offshore wind turbines installed in Europe},
  author={Bilgili, Mehmet and Alphan, Hakan and Ilhan, Akin},
  journal={Environmental Science and Pollution Research},
  volume={30},
  number={10},
  pages={27208--27226},
  year={2023},
  publisher={Springer}
}

\section*{Acknowledgements}

The authors would like to thank TotalEnergies for providing financial support for this project. We would also like to thank the Conselho Nacional de Desenvolvimento Cient{\'i}fico e Tecnol{\'o}gico (CNPq) and Coordena\c{c}\~{a}o de Aperfei\c{c}oamento de Pessoal de N{\'i}vel Superior (CAPES) for providing scholarships to the authors.

\section*{Author contributions statement}

\textbf{Conceptualization:} All authors.
\textbf{Methodology:} All authors.
\textbf{Data Curation:} L\'{\i}via Sancho; Mauricio S. Silva; Elisa Passos; Larissa F. R. Jacinto; Rebeca S. Lyra; Nilton O. Moraes; Carina S. Bock; Douglas M. Nehme; Raquel Toste; Carlos H. Beisl; Patr\'{\i}cia M. Silva; Adriano O. Vasconcelos; Rian C. Ferreira.
\textbf{Visualization:} Luan C. V. Silva; Patricia M. Silva; Larissa F. R. Jacinto; L\'{\i}via Sancho; Mauricio S. Silva; Elisa Passos; Adriano O. Vasconcelos.
\textbf{Writing -- Original Draft Preparation:} L\'{\i}via Sancho; Mauricio S. Silva; Elisa Passos; Luan C. V. Silva; Larissa F. R. Jacinto; Raquel Toste.
\textbf{Validation:} Luan C. V. Silva; Rodrigo S. Luna; Mauricio S. Silva; L\'{\i}via Sancho; Elisa Passos; Jacques Honigbaum; Fernando A. Rochinha; Alvaro L. G. A. Coutinho; Alexandre G. Evsukoff.
\textbf{Writing -- Review \& Editing:} Luan C. V. Silva; Mauricio S. Silva; L\'{\i}via Sancho; Elisa Passos; Alexandre G. Evsukoff; Luiz P. F. Assad.
\textbf{Supervision:} Fernando A. Rochinha; Luiz P. F. Assad; Alvaro L. G. A. Coutinho; Laura Bahiense; Alexandre G. Evsukoff.
All authors reviewed and approved the final manuscript.

\section*{Competing interests}

The authors declare no competing interests.

\end{document}